\documentstyle[preprint,epsfig,aps]{revtex}

\epsfverbosetrue

\newcommand{\beq}{\begin{equation}}
\newcommand{\enq}{\end{equation}}

\newcommand{\Si}[1]{\mbox{Si$_{#1}$}}

\begin{document}
\title{Chemical Reactions of Silicon Clusters}
\author{Mushti V. Ramakrishna and Jun Pan\footnote{The Department of Physics,
New York University, New York, NY 10003-6621.}}
\address{The Department of Chemistry, New York University,
New York, NY 10003-6621.}
\date{Submitted to \jcp, \today}
\maketitle

\begin{abstract}

Smalley and co-workers discovered that chemisorption reactivities of
silicon clusters vary over three orders of magnitude as a function of
cluster size.  In particular, they found that \Si{33}, \Si{39}, and
\Si{45} clusters are least reactive towards various reagents compared
to their immediate neighbors in size.  We explain these observations
based on our stuffed fullerene model.  This structural model consists
of bulk-like core of five atoms surrounded by fullerene-like surface.
Reconstruction of the ideal fullerene geometry gives rise to four-fold
coordinated crown atoms and $\pi$-bonded dimer pairs.  This model
yields unique structures for \Si{33}, \Si{39}, and \Si{45} clusters
without any dangling bonds and thus explains their lowest reactivity
towards chemisorption of closed shell reagents.  This model is also
consistent with the experimental finding of Jarrold and Constant that
silicon clusters undergo a transition from prolate to spherical shapes
at \Si{27}.  We justify our model based on an in depth analysis of
the differences between carbon and silicon chemistry and bonding
characteristics.  Using our model, we further explain why dissociative
chemisorption occurs on bulk surfaces while molecular chemisorption
occurs on cluster surfaces.  We also explain reagent specific
chemisorption reactivities observed experimentally based on the
electronic structures of the reagents.  Finally, experiments on
\Si{x}X$_y$ (X = B, Al, Ga, P, As, AlP, GaAs) are suggested as a means
of verifying the proposed model.  We predict that \Si{x}(AlP)$_y$ and
\Si{x}(GaAs)$_y$ $(x = 25, 31, 37; y = 4)$ clusters will be highly
inert and it may be possible to prepare macroscopic samples of these
alloy clusters through high temperature reactions.

\end{abstract}

\section{Introduction}

Chemical reactions on surfaces are at the heart of many industrial
processes.  For this reason, there are concerted experimental and
theoretical efforts aimed at gaining microscopic understanding of the
surfaces.  Despite these efforts we still lack fundamental insights
into how a surface promotes chemical reactions.  This problem is partly
due to the inadequacy of the available computational methods to solve
the electronic structure of complex systems with many degrees of
freedom.  On the other hand, this problem is also due to the inadequacy
of the experimental tools to perform controlled chemical reactions on
well defined surfaces.

Recently, a new trend has emerged to solve this problem.  This line of
attack, pioneered by Smalley and co-workers, involves studying chemical
reactions on the surfaces of size controlled and well-annealed
clusters.  Such studies are expected to yield a wealth of information
on the role of the size and composition of the surface in promoting
chemical reactions.  In this context, the experimental results obtained
by Smalley and co-workers on the chemisorption reactivities of
silicon clusters as a function of size are noteworthy
\cite{Elkind:87}.  These studies revealed that the reactivity rates for
ammonia (NH$_3$) chemisorption on cluster surfaces
are orders of magnitude smaller than
those found on bulk surfaces.  In addition, these reactivity rates vary
over three orders of magnitude as a function of cluster size, with 21,
25, 33, 39, and 45 atom clusters being particularly unreactive.
Clusters containing more than forty seven atoms do not display such
strong oscillations in reactivities as a function of the cluster size.
Similar results were obtained with methanol (CH$_3$OH), ethylene
(C$_2$H$_4$), and water (H$_2$O).  Only well annealed clusters
exhibited such pronounced oscillations in their reactivity.  But
reactions of silicon clusters with nitric oxide (NO) and oxygen (O$_2$)
did not reveal any oscillations in reactivity and all clusters reacted
readily with these reagents.  So chemical reactions of silicon clusters
seem sensitive to the cluster size, annealing, and reagent used.  In
initial experiments, H$_2$O seemed to behave more like NO and O$_2$
than like NH$_3$, CH$_3$OH, and C$_2$H$_4$.  However, later experiments
showed that even though H$_2$O was more reactive than ammonia it still
exhibited the same oscillatory pattern in reactivities as ammonia.

A simple explanation for these observations may be that some clusters
are much more stable than others and hence they are highly unreactive.
However, experimentally the abundance distribution of the silicon
clusters in the supersonic beam is a smoothly varying function of
cluster size, rather than being an oscillatory function
\cite{Elkind:87}.  Hence, oscillations in cluster reactivity are not
due to corresponding oscillations in their stability.

Alternative models based on structures of silicon clusters have been
proposed to explain the observed reactivity trends
\cite{Phillips:88,Jelski:88,Kaxiras:89,Patterson:90,Swift:91,Roth:94}.
However, none of these models explain all the experimental data.
Furthermore, the same model does not yield all cluster structures
consistently.  These models also do not satisfy the essential criterion
that the structures of the unreactive clusters should have zero
dangling bonds.  The effect of reagents on the chemical reactivity has
also not been explained.

In this Article we propose a consistent model that generates the
structures of the intermediate sized unreactive silicon clusters in a
systematic way.  The structures thus generated are unique for \Si{33},
\Si{39}, and \Si{45} clusters.  Furthermore, these clusters do not have
any dangling bonds and hence explains why these clusters are unreactive
for chemisorption.  Our model is consistent with the experimental
finding of Jarrold and Constant that at around \Si{27} the shapes of
silicon clusters undergo a transition from prolate to spherical shapes
\cite{Jarrold:91}.  We make an in depth analysis of the differences in
carbon and silicon chemistry and explain why the structures,
stabilities, and reactivities of silicon clusters are fundamentally
different from those of carbon clusters.  We explain why molecular
chemisorption takes place on cluster surfaces, whereas dissociative
chemisorption takes place on Si(111)-$(7 \times 7)$ surface.  We also
explain reagent specific chemisorption reactivities observed
experimentally.  Finally, we propose experiments on doped silicon
clusters that will test the validity of the proposed model and pave the
way for the synthesis of advanced cluster assembled materials through
high temperature reactions.

This paper is organized as follows: In Sec. II we present our stuffed
fullerene model  and in section III we analyze the differences between
carbon and silicon chemistry that justifies our model.  Using our
proposed cluster structures, we then explain in Sec. IV why
dissociative chemisorption occurs on bulk surfaces while molecular
chemisorption occurs on cluster surfaces.  In Sec. V we compare our
model with other models recently proposed to explain the experimentally
observed chemical reactivity pattern.  The reagent effect on
chemisorption reactivities is explained in Sec. VI  and in Sec. VII we
propose experiments on doped silicon clusters as a means of verifying
our model.  The results of this paper are summarized in Sec. VIII.

\section{Stuffed Fullerene Model}
Our structural model for silicon clusters consists of 1) a central atom
A, 2) four atoms B surrounding A in tetrahedral geometry, and 3)
fullerene surface.  The B atoms bind to twelve surface atoms, thus
rendering the B atoms also bulk-like with four-fold coordination and
tetrahedral geometry.  The surface then relaxes from its ideal
fullerene geometry, the same way the $(1 \times 1)$ bulk surfaces
relax.  This relaxation gives rise to crown atoms and dimers (CAD)
pattern on the surface.  The crown atoms are formally three-fold
coordinated and possess one dangling bond each.  The dimers are also
formally three-fold coordinated, but they eliminate their dangling
bonds through $\pi$ bonding.  The essential feature of this
construction is that the bulk-like core of five atoms (A + B) and
fullerene-like surface make these structures stable.  Fig. 1 presents a
graphical illustration of this stuffed fullerene model.  In principle,
this model is applicable to clusters containing more than twenty
atoms.

The proposed model yields structures with seventeen four-fold
co-ordinated atoms, four crown atoms, and the rest dimers.  The crown
atoms are called adatoms in surface science literature.
The word adatom connotes adsorbed atom, thereby making one think that
this is a weakly bonded atom.  In reality, this atom is strongly bonded
to the cluster.  Consequently, we prefer to call it the crown atom,
since it is at the center of three fused pentagons.

Unlike carbon, silicon does not form strong delocalized $\pi$ bonds
\cite{Brenner:91}.  Consequently, fullerene cage structures
\cite{Curl:91,Boo:92,Fowler:92} are energetically unfavorable for
silicon clusters \cite{Menon:94}.  Instead, intermediate sized silicon
clusters prefer $\sigma$-bonded network structures similar to the
diamond structure of bulk silicon \cite{Menon:94}.  The fullerene
geometry for the surface, consisting of interlocking pentagons and
hexagons, gives special stability to the surface atoms \cite{Boo:92}.
Furthermore, since delocalized $\pi$ bonding is not favorable in
silicon, we expect the surface atoms to relax from their ideal
fullerene geometry to allow for dimer formation through strong local
$\pi$ bonding.  Our model accounts for all these facts.

The 5-atom core (A + B) in our model has bulk-like geometry with one
atom in the center (A) bonded to four atoms (B) arranged in a perfect
tetrahedral symmetry.  There are numerous ways to orient the 5-atom
core inside the fullerene cage.  Furthermore, structural isomers exist
for fullerenes of any size \cite{Boo:92,Fowler:92}.  Thus we may use
this model to generate all possible structural isomers of odd numbered
intermediate sized clusters.  However, a particular orientation of the
5-atom pyramid and choosing the highest symmetry fullerene cages will
yield structures with the least number of dangling bonds.  We expect
such isomers to be most stable.

In the 20-60 atom size range only the 28- and 40-atom fullerene cages
belong to the $T_d$ point group (Table I) \cite{Boo:92}.  We generate
the \Si{33} and \Si{45} structures by stuffing the 5-atom core inside
these cages.  We orient the 5-atom pyramid in such a way that the
central atom A, the core atom B, and the crown atom C
lie on a line.  The crown atom is at the center of three fused
pentagons and it is surrounded by three other surface atoms D.
The D atoms now relax inwards to form the B-D bond.  The same type of
relaxation motion is necessary to form the $2 \times 1$ reconstruction
on the bulk Si(111) surface \cite{Cohen:84,Lannoo:91}.  The activation
barrier of $\approx$ 0.01 eV \cite{Cohen:84,Northrup:82} for this
relaxation is easily recovered by the formation of the B-D bond, whose
strength is 2.3 eV/bond \cite{Lannoo:91,Brenner:91}.  Consequently,
such a relaxation of fullerene surface is feasible even at 100 K.
Finally, the remaining surface atoms E readjust to form as
many dimers as possible.  Similar relaxation motion and dimer formation
occurs on Si(100) surfaces also \cite{Lannoo:91,Chadi:79}.  The dimers
are $\sigma$-bonded pair of atoms whose dangling bonds are saturated
through the formation of $\pi$ bonds.  Since the 5-atom core and the
fullerene cages have tetrahedral symmetry, the final \Si{33} and
\Si{45} structures will also have tetrahedral symmetry, if the
relaxation of the cage maintains this symmetry.

The \Si{39} structure is also generated in this way, starting from the
34-atom fullerene cage and stuffing 5-atom core inside it.  The final
\Si{39} structure has only C$_{3v}$ symmetry, because of the lower
symmetry of the fullerene cage \cite{Boo:92}.   The lower symmetry of
this structure gives rise to two different types of crown atoms.  The
crown atom along the C$_3$ axis is made of three fused hexagons,
whereas each of the other three crown atoms are made of three fused
pentagons.  Thus one crown atom in \Si{39} is in a slightly different
chemical environment than the other three.  In contrast, all the four
crown atoms in \Si{33} and \Si{45} are in chemically equivalent
environments because of the higher T$_d$ symmetry of their structures.
This is the only difference in the surface structure of these three
clusters.

The structures of \Si{33}, \Si{39}, and \Si{45} clusters thus generated
are displayed in Fig. 2.  The coordinates of these clusters are
tabulated in Table II.  These structures have four crown atoms, five
atoms in bulk-like environment, twelve four-fold coordinated surface
atoms not in bulk-like environment, and the rest as dimer pairs.  The
\Si{33} structure has six isolated dimers, \Si{45} has twelve dimers
forming four hexagons, while \Si{39} has nine dimers, three of which
form a hexagon.

We attempted to extend our model to reactive clusters also by
constructing \Si{35} and \Si{43} cluster structures.  However, we found
that these structures do not possess the characteristic CAD pattern
found in unreactive clusters.  This pattern, displayed in Fig. 3, is
responsible for eliminating the dangling bonds and only the unreactive
clusters seem to possess it.  The same pattern is found on Si(111)-$(7
\times 7)$ bulk surface also and scanning tunneling microscopy (STM)
\cite{Lieber:94} experiments of Wolkow and Avouris have revealed that
this is the least reactive site on this surface \cite{Wolkow:88}.  Our
model generates cluster structures with such a surface bonding
automatically.  The clusters that do not fit this model will have
larger number of dangling bonds and hence will be highly reactive.

As mentioned above, \Si{39} has a CAD pattern that is made of three
fused hexagons.   The bulk surfaces do not exhibit such a pattern and
hence we do not know at present the effect of this pattern on
reactivities.  However, the crown atom in this pattern also eliminates
dangling bonds and hence it is functionally equivalent to the CAD
pattern made of pentagons.  Consequently, we expect \Si{39} reactivity
trends to be similar to \Si{33} and \Si{45} clusters.

In principle, the proposed model can be used to construct \Si{25}
structure also.  However, \Si{20} cage is too small to accommodate five
atoms inside it without undue strain.  Consequently, we considered
building the \Si{25} cluster by inserting one atom inside a \Si{24}
cage.  However, such a structure will have large surface/volume ratio
and hence is likely to be unstable.  For these reasons, we presume that
\Si{25} will prefer a different geometry.  Indeed, the experiments of
Jarrold and Constant indicate that clusters smaller than \Si{27} do not
possess spherical shapes characteristic of larger clusters; instead
they seem to prefer prolate shapes \cite{Jarrold:91}.  Our inability to
generate a spherical structure for \Si{25} is consistent with this
experimental observation also.

We verified our model by constructing structures of all the clusters
discussed here using the ball and stick molecular modeling kits
\cite{Jones:94}.   These are rigid models allowing only tetrahedral
bond angles, whereas the proposed structures have non-tetrahedral
angles too.  Consequently, it was not possible to build stable
structures using these rigid modeling kits.  Nonetheless, we were able
to verify that only 33-, 39-, and 45-atom clusters possess the CAD
pattern on their surfaces.  We then hand calculated the cartesian
coordinates of these clusters from the tetrahedral geometry of the
5-atom core and the known geometries of the fullerene structures
\cite{Boo:92}, assuming that all bond lengths are approximately equal
to the bulk value of 2.35 \AA.  The structures thus generated are
displayed in Fig. 2.

We then relaxed these structures by carrying out molecular dynamics at
100 K using the Kaxiras-Pandey potential \cite{Kaxiras:88}.   The final
structures obtained from these simulations are nearly identical to the
initial ones.  This indicates that the proposed structures are locally
stable.  We could not do thorough relaxation of the proposed structures
through melting and recrystallization simulations, because this potential
is much more appropriate for bulk silicon than for clusters.

The structures displayed in Fig. 2 revealed that the crown atoms are
able to form a fourth bond to the core atoms B, thus rendering the B
atoms formally five-fold coordinated.  The B-C bond arises from the
back donation of the electrons from C to B and it weakens the
neighboring bonds through electronic repulsion.  We do not know the
strength of this back bond, but it is sufficiently strong to eliminate
the dangling bond on the crown atom and make these magic number
clusters unreactive.

The presence of five-fold coordinated atoms in these clusters seems
surprising at first.  However, this arises naturally from our model and
is consistent with the available calculations and experiments.
Electronic structure calculations of silicon clusters have shown that
clusters containing up to ten atoms exhibit multiple coordination with
a maximum coordination number of six \cite{Krishnan:85}.  Furthermore,
STM experiments on Si(111)-$(7 \times 7)$ surface have shown that the
atoms in the subsurface layer directly beneath the crown atoms are
five-fold coordinated \cite{Wolkow:88}.

The classical interatomic potentials available at present
\cite{Kaxiras:88,Stillinger:85,Biswas:85,Tersoff:86,Chelikowsky:89,Bolding:90}
may not be suitable for describing
unusual bonding patterns, such as the five-fold coordinated silicon
atoms found in these clusters.  Consequently, classical potentials
cannot be used to prove or disprove the validity of the proposed
structures.  Such a proof would have to come through {\em ab initio}
electronic structure calculations.  However, such calculations for
structural determination are extremely demanding computationally at
present for these clusters, even with the use of only modest size basis
sets.  Furthermore, such complicated calculations do not yield insights
into cluster properties, especially their reactivities.  Our simple
physically motivated model provides the conceptual framework for
understanding the nature of bonding in silicon clusters and explains
the experimental trends in reactivities.  This modeling effort at
understanding structures and chemical reactivities is similar in spirit
to the way the magic numbers of metal clusters are successfully
explained using the jellium model \cite{Cohen:90}.

\section{Analysis of Carbon and Silicon Chemistry}

The bonding characteristics of silicon differ in subtle but important
ways from that of carbon.  Much of carbon chemistry may be explained
based on sp, sp$^2$, and sp$^3$ hybridization of valence orbitals.
Such a simple hybridization scheme is possible because of the lack of
empty d-orbitals in carbon.  Furthermore, the small size of carbon does
not allow crowding of atoms around it due to steric repulsion between
these atoms.  For this reason, carbon always forms low-coordination
compounds, with the maximum coordination of four.  These two factors,
lack of empty d-orbitals and small size, constrain carbon to assume
simple sp$^n~ (n \le 3)$ hybridization, which automatically gives rise
to strong delocalized $\pi$ bonds and low coordination.

The behavior of silicon is exactly opposite to that of carbon.  Silicon
has empty d-orbitals that allows complicated hybridizations involving
s, p, and d orbitals.  Furthermore, it has a larger covalent radius
that permits crowding of several atoms around it without inducing large
steric repulsion.  Because of these two reasons, silicon structures
lower their energy through multiple bonding rather than through the
formation of delocalized $\pi$-bonded networks \cite{Menon:94}.

The consequences of these different bonding patterns of carbon and
silicon manifest themselves in their differing chemical and material
properties.  The following two lists illustrate some of these differences.
\begin{enumerate}
\item Graphite is the most stable form of carbon at room
temperature and atmospheric pressure \cite{Madelung:91}.
\item The diamond phase of carbon is metastable, its cohesive energy being
0.02 eV/atom less than that of the graphite phase \cite{Yin:83}.
\item Metallic (high-coordination) phases of carbon are not known
\cite{Madelung:91}.
\item Bulk C(111)-$(2 \times 1)$ reconstruction is stable \cite{Pate:86}.
\item Bulk C(111)-$(7 \times 7)$ is not observed \cite{Bokor:86}.
\item The maximum coordination of any stable carbon based material is four.
Five-fold coordinated carbon species, such as carbocations,
are unstable reaction intermediates.  CH$_5^+$ is an important example
of a carbocation, one that is used in chemical ionization mass spectrometry.
\item Carbon clusters form stable chain, ring, and cage structures
\cite{Krishnan:87,Andreoni:90,Parasuk:91,Menon:93-2}.
\end{enumerate}
In sharp contrast, the behavior of silicon is very different.
\begin{enumerate}
\item Diamond is the most stable phase of silicon at room temperature
and atmospheric pressure \cite{Madelung:91}.
\item Graphite form of silicon is unstable, and has never been
experimentally prepared \cite{Yin:82}.
\item The metallic (high-coordination) $\beta$-tin phase of silicon is
0.2 eV/atom less stable than the diamond phase \cite{Yin:82}.  Also,
liquid silicon is a metal.
\item Bulk Si(111)-$(2 \times 1)$ reconstruction is metastable
\cite{Lannoo:91}.
\item Bulk Si(111)-$(7 \times 7)$ reconstruction is stable \cite{Lannoo:91}.
\item Silicon clusters do not form chain, ring, and cage structures.
Instead they prefer $\sigma$-bonded multiply coordinated
network structures with a maximum coordination of six
\cite{Krishnan:85,Tomanek:86,Menon:93-1}.
\end{enumerate}
These examples illustrate how subtle differences in bonding
characteristics give rise to vastly different crystal structures
and surface reconstructions.  Indeed, these examples are manifestations
of the effect of local electronic structure of individual atoms on the
material properties at macroscopic level.  Our model of silicon
clusters accounts for all these facts by focusing on structures that
are able to form maximum number of $\sigma$ bonds and eliminate their
surface dangling bonds through local $\pi$ bonding.

The stabilities of clusters is related to their cohesive energies:
larger the cohesive energy greater its stability.  If certain clusters
are more stable than their neighbors in size then we would see
oscillations in the mass spectra of cluster abundance in the supersonic
cluster beam.  This is the case for carbon clusters where C$_{60}$
dominates all others because of its special stability.  However,
silicon clusters do not show any such special abundance in the 20-60
atom size range, indicative that in this size range the cluster
stability is a smoothly varying function of size.  In particular, the
\Si{33}, \Si{39}, and \Si{45} clusters are not more stable than their
neighbors; only that they are much less reactive than their neighbors.

At first this behavior, being different from that of carbon, appeared
to be contradictory.  However, that is not the case.  In the case of
silicon, these two observations only reinforce the notion that the
$\sigma$-bonding is the most important factor determining the cluster
stability and $\pi$-bonds in silicon are weak.  However, these
$\pi$-bonds are important in eliminating the surface dangling bonds and
thus make the clusters less reactive.  In other words, the cluster
stability is mostly determined by $\sigma$-bonds whereas the cluster
reactivity is sensitive to the dangling bonds.  This explains why oscillations
are seen only in the silicon cluster reactivities but not in their
abundance mass spectra.

In the case of carbon, $\pi$-bonds are strong and they play an
important role in cluster stability as well as reactivity.  Certain
cluster sizes that allow delocalized $\pi$-bonded networks to be formed
are more stable than their neighbors, while also being relatively more
inert.  The inertness of these clusters arises from the strength of the
$\pi$-bonds.  For this reason, most stable carbon clusters are also
least reactive.

Since the study of carbon clusters preceded that of silicon clusters
chronologically, it was expected that silicon clusters would behave the
same way as carbon clusters.  When that did not happen, it gave rise to
some confusion in the literature and even to speculation that perhaps
the experiments of Smalley and co-workers are wrong.  This speculation
was further fueled by Martin Jarrold's experiments which did not show
strong oscillations in the reactivities of silicon clusters in the
30-50 atom size regime \cite{Jarrold:89}.  At present there is considerable
evidence from Smalley's experiments that the clusters in Jarrold's
experiments were not well annealed \cite{Elkind:87}.
This point is still being debated,
but our own theoretical study supports Smalley's experiments.

\section{Chemisorption on Bulk vs. Cluster Surfaces}

Ammonia is known to dissociatively chemisorb on the bulk Si(111)-$(7
\times 7)$ surface \cite{Kubler:87,Tanaka:87,Bozso:88}.  On the other
hand, ammonia chemisorption on silicon cluster surfaces seems to be
occurring without any loss of hydrogen atoms \cite{Elkind:87}.  Since,
dissociative chemisorption would have more likely resulted in the loss
of hydrogen, we presume that ammonia chemisorption on silicon cluster
surfaces is molecular.  In addition, the bulk silicon surface seems to
be orders of magnitude more reactive than any of the cluster surfaces.

We can explain both these differences in cluster and bulk surface
reactivities using our stuffed fullerene model.  On the bulk
Si(111)-$(7 \times 7)$ surface there are six rest atoms and 12 crown
atoms per unit cell \cite{Takayanagi:85}.  The rest atom and crown atom
structures are depicted in Fig. 4.
{}From STM studies we learn that dissociative chemisorption on
Si(111)-$(7 \times 7)$ seems to be related to the proximity of the rest
atom to a crown atom \cite{Wolkow:88}.  The rest atom is highly
reactive, because it has a dangling bond.  The energy released from the
reaction between the rest atom and ammonia dissociates ammonia into
NH$_2$ and H.   One fragment binds to the rest atom and the other binds
to the nearby crown atom.  It is not known at present whether these
fragments bind specifically to one site and not the other.  The
detailed mechanism of this reaction is not known either.  However, such
a dissociative chemisorption is not possible if rest atoms are not
present, as in our proposed cluster structures.  Furthermore, absence
of rest atoms will make these clusters much less reactive than bulk
surfaces, which have six rest atoms per unit cell.  Thus our model is
consistent with the experimental observations of Smalley and co-workers
and also of the reactivity experiments on bulk Si(111)-$(7 \times 7)$
surface.

The bulk Si(111)-$(7 \times 7)$ surface has two types crown atoms, the
corner and the center crown atoms.  The corner crown atom is surrounded
by two pairs of dimers whereas the center crown atom is surrounded by
one dimer.  The corner crown atoms are even less reactive than the
center crown atoms \cite{Wolkow:88}.  The crown atoms on the proposed
cluster surfaces are more like the corner crown atoms because they are
surrounded by three pairs of dimers.  The reactivity seems to decrease
with increasing number of dimers around the crown atom site
\cite{Wolkow:88}.  This explains why our proposed cluster structures
exhibit lowest reactivity.

\section{Previously proposed models}

Several alternative structural models
\cite{Phillips:88,Jelski:88,Kaxiras:89,Patterson:90,Swift:91,Roth:94}
have been proposed in recent years to explain the chemisorption
reactivity data of Smalley and co-workers \cite{Elkind:87}.
Unfortunately, none of these proposed models give a consistent
explanation of all the experimental data.  Nonetheless, we give a brief
critique of these models and their strengths and weaknesses.

Kaxiras has proposed structures of Si$_{33}$ and Si$_{45}$ clusters
based on the reconstructed 7 $\times$ 7 and 2 $\times$ 1 surfaces of
bulk Si(111), respectively \cite{Kaxiras:89}.  The hypothesis here is
that the interior atoms are most bulk-like while the surface atoms are
reconstructed bulk surface-like.  As the cluster size is increased they
smoothly approach the bulk limit with diamond structure in the interior
and reconstructed surfaces on the outside.  Consequently, this model
exhibits the correct limiting behavior as the cluster size $N
\longrightarrow \infty$.  Based on this model, Kaxiras constructed
structures of \Si{33} and \Si{45} clusters.  He then argued that adding
or subtracting one atom from these magic clusters would create defects
on the surface that would be highly reactive.  Kaxiras thus explains
why certain clusters are less reactive than others.  However, the exact
nature of the defect states on the surface is not specified.
Furthermore, the model itself is defective on several counts: 1) The
bulk surfaces are even more reactive than the clusters; 2) The surface
of \Si{45} is the metastable 2 $\times$ 1 surface rather than the
globally stable 7 $\times$ 7 surface \cite{Lannoo:91} and the
Si(111)-$(2 \times 1)$ surface is much more reactive than the $7 \times
7$ surface;  3) The model may not uniquely apply just to the magic
clusters.  With some effort one may be able to construct reactive
silicon clusters that obey the Kaxiras model.

The \Si{45} structure of Kaxiras has forty dangling bonds, which make
it highly reactive, contrary to the experiments.  To overcome this
discrepancy between experiment and theory, Kaxiras has postulated that
the dangling bonds on \Si{45} form $\pi$-bonded chains, analogous to
the bulk Si(111)-$(2 \times 1)$ surface \cite{Pandey:81}.  However,
silicon favors strong local $\pi$ bonds over delocalized $\pi$-bonded
chains.  This is the reason why silicon does not form fullerene cages
and the Si(111)-$(2 \times 1)$ reconstruction, involving $\pi$-bonded
chains \cite{Pandey:81}, is metastable with respect to the locally
$\pi$-bonded $7 \times 7$ reconstruction \cite{Lannoo:91}.
Consequently, the \Si{45} structure proposed by Kaxiras is a metastable
and highly reactive structure.  Indeed, Jelski and co-workers disputed
the Kaxiras model of \Si{45} by constructing alternative structures
that are lower in energy, but do not possess any of the features of the
reconstructed bulk surfaces \cite{Swift:91}.

Our structure for \Si{33} is identical to that proposed by Kaxiras
\cite{Kaxiras:89} and Patterson and Messmer \cite{Patterson:90}.  This
structure has been shown to be locally stable through density
functional calculations \cite{Feldman:91}.  But our \Si{45} structure
is different from that of Kaxiras \cite{Kaxiras:89}.  However, we can
generate the \Si{45} structure of Kaxiras by stuffing one atom inside a
44-atom fullerene cage and allowing for the reconstruction of the
fullerene surface.  Thus our model is very general and the Kaxiras
model is a subset of our model.

Jelski proposed a very innovative structure for \Si{45}
\cite{Swift:91}.  This structure is most bulk-like of all the proposed
structures so far and it is stable.  However, we do not know if this
structure should be least reactive.  Furthermore, Jelski proposed only
\Si{45} structure until now and we do not know if it is possible to
construct similar structures for \Si{33} and \Si{39}.

More recently, Andreoni and co-workers have determined the structures
of \Si{45} clusters using their Car-Parrinello molecular dynamics
technique \cite{Roth:94}.  They found that the surface of \Si{45} is
fullerene-like, but its interior contains two or more atoms, which are
not in a bulk-like environment.  This is a very surprising result,
since it indicates that the bulk-like structure has not set in even in
these intermediate sized clusters.  In the absence of full details of
the search algorithm employed in these calculations, it is difficult to
say whether are not the structure obtained in these simulations is the
global minimum.  Furthermore, the calculations do not explain the
experimental reactivity trends observed by Smalley and co-workers.
However, what emerges from this work is that the surface of silicon
clusters is fullerene-like, in agreement with our proposed stuffed
fullerene model developed independently.

\section{Reagent Effect on Chemical Reactivities}

The reactivity patterns of NO and O$_2$ are different from those found
for NH$_3$, CH$_3$OH, C$_2$H$_4$, and H$_2$O \cite{Elkind:87}.  This
reagent effect on chemical reactivities may be explained based on the
ground state electronic structures of the reagents.  NH$_3$, CH$_3$OH,
C$_2$H$_4$, and H$_2$O in their ground states have closed shell
electronic structure with all electrons paired.  On the other hand, NO
and O$_2$ in their ground states are $^2\Pi_g$ and $^3\Sigma_g^-$,
possessing one and two unpaired electrons, respectively
\cite{Herzberg:50}.  Consequently, NH$_3$, CH$_3$OH, C$_2$H$_4$, and
H$_2$O can chemisorb only at those sites where excess electron density
is present due to dangling bonds.  Such a selectivity gives rise to
highly oscillatory pattern in the reactivities, because the number of
dangling bonds varies as a function of cluster size.  The magic number
clusters are unreactive because they do not possess any dangling
bonds.  On the other hand, NO $(^2\Pi_g)$ and O$_2$ $(^3\Sigma_g^-)$
can bind anywhere, since they have their own dangling bonds that can
instigate reaction anywhere on the cluster surface.  Consequently, NO
and O$_2$ readily react with all clusters and do not display the
oscillatory pattern in their chemical reactivities.  This explains the
reagent specific chemisorption reactivities observed experimentally
\cite{Elkind:87}.

The magic number clusters are less reactive than others but they are
not completely inert towards the closed shell reagents
\cite{Elkind:87}.  These clusters are most inert towards C$_2$H$_4$
than towards NH$_3$, CH$_3$OH, and  H$_2$O.  This is because NH$_3$,
CH$_3$OH, and H$_2$O have lone pairs on either nitrogen or oxygen and
these lone pairs can instigate reaction on the cluster surface.
C$_2$H$_4$ does not have any lone pairs and hence the magic number
clusters are highly inert towards this reagent.  The electronic
structure of reagents can thus explain even subtle differences in the
reactivities of magic number clusters towards a group of related
reagents.

\section{Doped silicon clusters}

It is interesting to speculate on how one may confirm the proposed
structures experimentally.  From the reactivities on bulk surfaces, we
know that Si(111)-$(7 \times 7)$ surfaces doped with group III and V
elements are much less reactive than the pure surfaces
\cite{Wolkow:88}.  This is because these dopants eliminate the dangling
bonds much more effectively and thus reduce the surface reactivity.
Since the surfaces of proposed silicon clusters are very similar to the
bulk Si(111)-$(7 \times 7)$ surface, it is reasonable to expect that
alloy clusters of silicon with group III and V elements will be much
less reactive than the pure clusters.  Furthermore, the reactivity
varies with the number of dopant atoms and minimal reactivity occurs
when all the dangling bonds are fully saturated.

For example, the reactivities of annealed Si$_x$X$_y (x + y = 33, 39,
45;$ X = B, Al, Ga, P, As) clusters vary as y is varied.  From the
experimental studies on Si(111)-$(7 \times 7)$ surfaces we know that X
= (B, P, As) dopants occupy the five-fold coordinated core atom B sites
in Fig. 2 \cite{Wolkow:88}.  There are four such sites and hence we
expect the reactivity to be a minimum when $y = 4$.  Likewise, the
dopants X = (Al, Ga) occupy crown atom sites C \cite{Wolkow:88}.  Once
again, there are four such sites and hence the reactivity will be a
minimum when $y = 4$, if our proposed model of silicon clusters is
correct.

Si$_x$(AlP)$_y$ and Si$_x$(GaAs)$_y$ alloys are probably the most
interesting systems to study.  AlP is isoelectronic to Si$_2$ and the
sizes of Al and P are close to that of Si.  Furthermore, bulk AlP
crystal structure is identical to that of silicon and its lattice
constant is only 0.31 \AA~ larger than that of bulk silicon
\cite{Madelung:91}.  The
calculated structures of (AlP)$_2$ and (AlP)$_3$ are nearly the same as
\Si{4} and \Si{6} cluster structures \cite{Allaham:92}.  Hence, we
expect the larger Si$_x$(AlP)$_y$ clusters to take the same cluster
structures as Si$_{x+2y}$ clusters.  In addition, we expect the
reactivities of Si$_x$(AlP)$_y$ clusters also to be a minimum when $x +
2y = 33, 39, 45$ and $y = 4$.  The four AlP molecules in these least
reactive clusters will occupy sites that eliminate all dangling bonds.
Aluminum will occupy crown atom C sites while P will occupy the core
atom B sites.  Such clusters will be much less reactive than pure Si
clusters and it may even be possible to synthesize them in macroscopic
samples.

Even though GaAs is not isoelectronic to Si$_2$, it is quite non-polar,
its bulk lattice structure is identical to that of silicon, and its
lattice constant is only 0.22 \AA~ larger than that of bulk silicon
\cite{Madelung:91}.  Consequently, it is possible that the structures
of Si$_x$(GaAs)$_y$ clusters will be the same as those of pure silicon
clusters.  In these alloy clusters, we expect As to occupy crown atom C
sites and Ga to occupy core atom B sites.  Four molecules of GaAs will
then eliminate all dangling bonds.  Consequently, we predict that
Si$_x$(GaAs)$_y$ $(x + 2y = 33, 39, 45; y = 4)$ clusters to be highly
inert.  It may also be possible to isolate such clusters in macroscopic
samples through high temperature reactions, like fullerenes.

{}From this speculation it is clear that alloy clusters may pave the way
towards the synthesis of novel advanced materials of interest to
optical, electronic, and communications industry.

\section{Summary}

In summary, we propose a novel structural model to explain the
experimental data of Smalley and co-workers on the chemisorption
reactivities of silicon clusters towards various reagents
\cite{Elkind:87}.  In principle, this model is applicable to silicon
clusters containing more than twenty atoms.  It consists of bulk-like
core of five atoms stuffed inside reconstructed fullerene cages.  The
resulting structures of \Si{33}, \Si{39}, and \Si{45} are unique, have
maximum number of four-fold coordinated atoms, minimum number of
surface atoms, and zero dangling bonds.  This model does not yield such
unique structures for other intermediate sized clusters and hence they
will have larger number of dangling bonds.  This explains why \Si{33},
\Si{39}, and \Si{45} clusters are least reactive towards reagents with
closed shell electronic structure, such as ammonia, methanol, ethylene,
and water \cite{Elkind:87}.  Our model also indicates that \Si{25}
cluster cannot be formed in a spherical shape.  This result is
consistent with the experimental finding of Jarrold and Constant that
at around \Si{27} silicon clusters undergo a shape transition from
prolate to spherical shapes \cite{Jarrold:91}.  We justify our model
through an in depth analysis of the similarities and differences of
carbon and silicon bonding and chemistry.

Chemisorption is sensitive to the local electronic structure and
chemical bonding on the surface.  On bulk Si(111)-($7 \times 7$)
surface, chemisorption occurs preferentially on rest atom sites and
less on crown atom sites \cite{Wolkow:88}.  Absence of rest atoms in
silicon clusters is the main reason why these clusters are much less
reactive than bulk silicon.  Furthermore, dissociative chemisorption of
ammonia on bulk Si(111)-($7 \times 7$) surface is due to the close
proximity of the rest atom and crown atom sites \cite{Wolkow:88}.
Absence of such a configuration on silicon clusters explains why
ammonia is unable to dissociate on the cluster surfaces; instead it
chemisorbs molecularly.

Two distinct patterns of chemisorption reactivities observed
experimentally are explained based on the electronic structures of the
reagents.  The reactivities of closed shell reagents are much more
sensitive to the cluster size than those of open shell reagents.  This
is because the closed shell reagents chemisorb on those sites that have
excess electron density due to dangling bonds.  Such a selectivity
gives rise to highly oscillatory chemisorption reactivities, since the
number of dangling bonds varies with cluster size.  The open shell
reagents are not so selective because they carry the dangling bonds
necessary to instigate reaction anywhere on the cluster surface.
Hence, silicon clusters of all sizes readily react with the open shell
reagents.  This explains why only the closed shell reagents are
sensitive to the cluster size, structure, and annealing and exhibit the
highly oscillatory pattern in reactivities as a function of the cluster
size.

Finally, we propose that experiments on doped silicon clusters may
yield further insights into the structures of pure clusters and the
reasons for their overall low chemical reactivities.  For example,
\Si{x}X$_y~ (x + y = 33, 39, 45; y = 4;$ X = B, Al, Ga, P, As, AlP,
GaAs) alloy clusters may be substantially inert because the dopants
saturate the dangling bonds on the cluster surface much more
effectively.  \Si{x}(AlP)$_4$ and \Si{x}(GaAs)$_4$ $(x = 25, 31, 37)$
clusters in particular may be so inert that it may be possible to
synthesize macroscopic samples of these clusters through high
temperature reactions.  Experiments on such alloy clusters are needed
to fully understand the microscopic origins of chemical reactivity
trends in the pure as well as alloy clusters and synthesize new
advanced cluster based materials \cite{Siegel:93}.  Simultaneously, STM
experiments on Si(111)-$(2 \times 1)$ and Si(111)-$(7 \times 7)$
surfaces using other closed shell reagents such as ethylene are needed
to fully understand elementary steps of chemisorption on microscopic
cluster and macroscopic bulk surfaces.

\section{Acknowledgments}
This research is supported by the New York University Research
Challenge Fund and the Donors of The Petroleum Research Fund (ACS-PRF
\# 26488-G), administered by the American Chemical Society.  The
graphics presented in this paper are all generated using XMol program
from the Research Equipment Inc. and the University of Minnesota
Supercomputer Center.

\begin{table}
\caption{The symmetry point groups of fullerenes of size N = 20-60.
Each fullerene may exist in several different isomeric forms.  The point
groups listed below are for the highest symmetry isomers.  }
\begin{center}
\begin{tabular}{cc}
N               &  Point group \\ \hline
20, 60          &  I$_h$    \\
28, 40          &  T$_d$    \\
44, 52          &  T      \\
36              &  D$_{6h}$ \\
24              &  D$_{6d}$ \\
30, 50          &  D$_{5h}$ \\
26              &  D$_{3h}$ \\
32, 56          &  D$_{3d}$ \\
38, 42, 48, 54  &  D$_3$    \\
34, 58          &  C$_3$    \\
\end{tabular}
\end{center}
\end{table}

\begin{table}
\caption{The cartesian coordinates of the proposed \Si{33} cluster structure.}
\begin{center}
\begin{tabular}{cccc}
Atom no.  &  X (\AA)  &  Y (\AA)  &  Z (\AA)  \\  \hline
 1  & -0.012  & -0.025  & -0.008 \\
 2  & -0.563  & -1.151  &  1.980 \\
 3  & -0.028  &  2.293  &  0.374 \\
 4  & -1.586  & -0.555  & -1.670 \\
 5  &  2.128  & -0.688  & -0.716 \\
 6  & -0.975  & -1.992  &  3.467 \\
 7  &  0.964  & -0.693  &  3.748 \\
 8  & -2.717  & -0.562  &  2.802 \\
 9  & -0.581  & -3.516  &  1.722 \\
10  &  0.419  &  1.539  &  3.257 \\
11  & -1.856  &  1.620  &  2.673 \\
12  & -3.400  & -1.202  &  0.647 \\
13  & -2.080  & -3.028  & -0.021 \\
14  &  1.601  & -3.159  &  0.924 \\
15  &  2.555  & -1.415  &  2.176 \\
16  &  1.531  &  2.960  &  2.044 \\
17  & -2.150  &  3.091  &  1.099 \\
18  & -3.803  &  0.069  & -1.069 \\
19  & -1.666  & -2.885  & -2.150 \\
20  &  2.273  & -3.025  & -1.138 \\
21  &  3.818  & -0.202  &  0.888 \\
22  & -0.040  &  4.027  &  0.660 \\
23  & -2.764  & -0.952  & -2.913 \\
24  &  3.729  & -1.183  & -1.246 \\
25  &  0.612  & -2.682  & -2.806 \\
26  &  3.219  &  2.084  &  0.615 \\
27  & -2.996  &  2.305  & -0.981 \\
28  &  0.517  &  3.549  & -1.572 \\
29  &  2.592  &  2.449  & -1.621 \\
30  & -1.136  &  0.528  & -3.741 \\
31  & -1.348  &  2.589  & -2.632 \\
32  &  2.804  &  0.387  & -2.729 \\
33  &  0.939  & -0.573  & -3.789
\end{tabular}
\end{center}
\end{table}

\begin{table}
\caption{The cartesian coordinates
of the proposed \Si{39} cluster structure.}
\begin{center}
\begin{tabular}{cccc}
Atom no.  &  X (\AA)  &  Y (\AA)  &  Z (\AA)  \\  \hline
 1  &  0.006  &  0.656  &  3.533 \\
 2  &  0.028  &  3.005  &  3.573 \\
 3  &  1.376  & -0.170  &  5.254 \\
 4  &  0.813  & -0.099  &  1.459 \\
 5  & -2.193  & -0.113  &  3.845 \\
 6  &  0.313  & -3.564  &  5.785 \\
 7  &  2.153  & -3.556  &  4.323 \\
 8  &  1.808  & -3.512  &  1.999 \\
 9  & -0.378  & -3.477  &  1.136 \\
10  & -2.218  & -3.486  &  2.598 \\
11  & -1.873  & -3.529  &  4.922 \\
12  &  0.743  & -2.020  &  7.503 \\
13  &  3.721  & -2.007  &  5.139 \\
14  &  3.080  & -1.926  &  0.820 \\
15  & -0.457  & -1.870  & -0.576 \\
16  & -3.877  & -1.885  &  2.140 \\
17  & -3.318  & -1.955  &  5.900 \\
18  &  2.850  & -1.058  &  7.104 \\
19  &  1.680  & -0.911  & -0.771 \\
20  & -4.557  & -0.939  &  4.181 \\
21  &  2.538  &  1.240  &  6.725 \\
22  &  1.504  &  1.370  & -0.234 \\
23  & -4.006  &  1.345  &  4.141 \\
24  &  0.375  &  2.093  &  7.062 \\
25  &  3.352  &  2.106  &  4.698 \\
26  &  2.877  &  2.166  &  1.499 \\
27  & -0.659  &  2.223  &  0.103 \\
28  & -3.192  &  2.211  &  2.114 \\
29  & -2.634  &  2.141  &  5.875 \\
30  & -0.773  &  3.555  &  5.625 \\
31  &  2.204  &  3.568  &  3.260 \\
32  & -1.332  &  3.625  &  1.864 \\
33  &  0.041  &  4.421  &  3.598 \\
34  & -3.506  &  0.135  &  0.997 \\
35  & -1.665  &  0.143  & -0.465 \\
36  & -0.434  & -0.011  &  7.822 \\
37  & -2.620  &  0.023  &  6.959 \\
38  &  3.942  &  0.053  &  1.749 \\
39  &  4.287  &  0.010  &  4.073
\end{tabular}
\end{center}
\end{table}

\begin{table}
\caption{The cartesian coordinates
of the proposed \Si{45} cluster structure.}
\begin{center}
\begin{tabular}{cccc}
Atom no.  &  X (\AA)  &  Y (\AA)  &  Z (\AA)  \\  \hline
 1  &  0.000  &  0.000  &  0.000 \\
 2  & -2.107  & -0.831  &  0.625 \\
 3  & -0.014  &  2.346  &  0.128 \\
 4  &  0.482  & -0.662  & -2.203 \\
 5  &  1.640  & -0.853  &  1.450 \\
 6  & -4.815  & -1.900  &  1.427 \\
 7  & -4.780  &  0.464  &  1.540 \\
 8  & -4.284  & -2.541  & -0.788 \\
 9  & -3.127  & -2.733  &  2.861 \\
10  & -3.067  &  1.074  &  3.026 \\
11  & -4.215  &  1.264  & -0.594 \\
12  & -3.911  & -0.578  & -2.021 \\
13  & -2.272  & -3.748  & -0.710 \\
14  & -1.562  & -3.866  &  1.526 \\
15  & -2.054  & -0.885  &  3.836 \\
16  & -1.785  &  3.020  &  2.722 \\
17  & -2.933  &  3.210  & -0.898 \\
18  & -2.326  & -0.474  & -3.753 \\
19  & -0.686  & -3.644  & -2.442 \\
20  &  0.732  & -3.879  &  2.032 \\
21  &  0.241  & -0.899  &  4.342 \\
22  & -2.121  &  4.476  &  0.907 \\
23  & -1.000  & -2.318  & -4.356 \\
24  &  1.616  & -2.751  &  3.895 \\
25  &  2.318  & -3.775  &  0.301 \\
26  &  1.608  & -3.658  & -1.936 \\
27  &  0.509  &  3.007  &  3.227 \\
28  &  1.523  &  1.047  &  4.037 \\
29  & -1.044  &  1.472  & -4.057 \\
30  & -1.348  &  3.314  & -2.630 \\
31  & -0.033  &  5.362  &  0.292 \\
32  &  1.100  & -1.512  & -5.034 \\
33  &  3.748  & -1.950  &  3.314 \\
34  &  1.593  &  4.454  &  1.726 \\
35  &  0.445  &  4.644  & -1.894 \\
36  &  1.073  &  0.831  & -4.849 \\
37  &  2.713  & -2.340  & -3.538 \\
38  &  4.182  & -2.583  &  1.093 \\
39  &  3.690  &  0.397  &  3.403 \\
40  &  3.075  &  3.175  &  0.426 \\
41  &  2.365  &  3.293  & -1.811 \\
42  &  2.669  &  1.451  & -3.239 \\
43  &  3.682  & -0.509  & -2.429 \\
44  &  4.392  & -0.626  & -0.191 \\
45  &  4.088  &  1.215  &  1.236
\end{tabular}
\end{center}
\end{table}

\begin{figure}
\caption{The 5-atom bulk-like core (left) + the 28-atom fullerene cage
(middle) = the proposed structure for the \Si{33} cluster (right).
This is the graphical illustration of the stuffed fullerene model
proposed in this paper.  }

\end{figure}

\begin{figure}
\caption{Structures of the least reactive  clusters a) \Si{33}, b)
\Si{39}, and c) \Si{45} obtained using the proposed stuffed fullerene
model.   These clusters do not possess any dangling bonds and hence are
least reactive towards reagents with closed shell electronic structure,
such as ammonia, methanol, ethylene, and water.  Representative atoms
in different chemical environments are labeled from A to E.  }

\end{figure}

\begin{figure}
\caption{The crown atom and dimer (CAD) structure consisting of three
fused pentagons.  This structure is the essential building block on the
surfaces of unreactive silicon clusters.  B is the core atom, C is the
crown atom, and E atoms are part of the dimers.  This structure is also
found on the bulk Si(111)-($7 \times 7$) surface.  The crown atom is
called the adatom in surface science literature.}

\end{figure}

\begin{figure}
\caption{The rest atom (R) and crown atom (C) structures.  The rest
atom has a dangling bond and hence is highly reactive.  The crown atom
saturates its dangling bond through back donation of the electrons to
the core atom B and hence is much less reactive than the rest atom.
The strength of the B-C back bond is not known at present, but it may
be weaker than the normal Si-Si bond due to the electronic repulsion
from the neighboring bonds.  On the bulk Si(111)-$(7 \times 7)$ surface
both R and C type of atoms are found but in our model of unreactive
silicon clusters only the crown atoms are found.  This explains low
reactivity of these clusters towards reagents with closed shell
electronic structure.  The absence of rest atoms in these cluster
structures also explains why dissociative chemisorption does not take
place on cluster surfaces, whereas this is the primary mechanism of
chemisorption on the bulk Si(111)-$(7 \times 7)$ surface.  }

\end{figure}

\epsfig{file=Fig1.ps}
\epsfig{file=Si33.ps}
\epsfig{file=Si39.ps}
\epsfig{file=Si45.ps}
\epsfig{file=Si11.ps}
\epsfig{file=Si9.ps}

\end{document}